\documentstyle[12pt,BoxedEPS]{article}
\font\sf=cmss10                    
\SetRokickiEPSFSpecial 
\HideDisplacementBoxes

\def\Figure#1#2{
\begin{figure} $$\BoxedEPSF{#1}$$
\caption {\sf #2}
\end{figure}}

\def\V{\cal V}
\def\del{\partial}
\def\ra{\rightarrow}
\def\delrho{\delta \: \!\! \rho}

\begin{document}

{\flushright{\small MIT-CTP-2667\\
hep-th/9708084\\}}

\vspace{1.0in}
\begin{center}

{\large MINIMAL AREA NONORIENTABLE STRING DIAGRAMS\\}
\vspace{0.8in}
{\large Oliver DeWolfe}\footnote{\noindent email: odewolfe@ctp.mit.edu \\
\hspace*{.15in} Work supported by the U.S.\ Department of Energy under contract \#DE-FC02-94ER40818.}
\vskip 0.2cm
{{\it Center for Theoretical Physics,\\
Laboratory for Nuclear Science,\\
Department of Physics\\
Massachusetts Institute of Technology\\
Cambridge, Massachusetts 02139, U.S.A.}}

\end{center}

\vspace{0.5in}

\begin{abstract} 
We use minimal area metrics to generate all  nonorientable string diagrams.
The surfaces in unoriented string theory have nontrivial open curves and
nontrivial closed curves whose neighborhoods are  either annuli or M\"{o}bius
strips.  We define a minimal area problem by imposing length conditions on 
open curves and on annular closed curves only.  We verify that the minimal area
conditions are respected by the sewing operations.  The natural objects that
satisfy recursion relations involving the propagator, which performs both
orientable and nonorientable sewing, are classes of moduli spaces grouped by
Euler characteristic.

\end{abstract}

\newpage

\section{Introduction}

In calculating perturbative string amplitudes one must integrate suitable
forms over moduli spaces of Riemann surfaces. For this purpose one requires a
concrete presentation of each surface in the moduli space, a string diagram,
and a rule to build all the string diagrams in a given moduli space.
A natural way to achieve this was demonstrated by Zwiebach~\cite{zwiebach}
using minimal area metrics.  The uniqueness of the minimal area metric is used 
to ascertain that each surface is produced once and only once.  One can
show that whenever surfaces with appropriate minimal area metrics are
sewn together, the resulting surface also has a minimal area metric --- the
so-called sewing theorem.  Using this technique all diagrams for a string
amplitude can be generated in a natural way.

The purpose of this paper is to extend this work to the previously-unexamined
case of surfaces which are not oriented.  This case is only natural to
consider, since Type I string theory is only consistent with gauge group 
SO(32), whose fundamental representation is real and which therefore
gives rise to nonorientable diagrams~\cite{gsw}.  We shall review nonorientable
surfaces, then proceed to extend the minimal area problem to include them.
We shall see that, while nontrivial open curves are no different from the
orientable case, nontrivial closed curves have neighborhoods that are either
annuli or M\"{o}bius strips, and the minimal area problem imposes length
conditions on the annular closed curves and open curves only.  It
is satisfying that length conditions are thus placed only on curves that
correspond to external string states.  It is additionally interesting that
M\"{o}bius curves are forced to have lengths no less than the shortest
{\it open} curves.

We also work through two explicit examples of covering moduli spaces, and
finally prove the sewing theorem and establish the nonorientable sewing
recursion relations.  For a detailed examination of the oriented
open-closed case, see~\cite{revisited}.  Constraints on conformal field
theories existing on nonorientable surfaces can be found
in~\cite{lewellen,fioravanti}.

\section{Review of Nonorientable Surfaces}

As is well-known, generic orientable surfaces in two dimensions can be
characterized topologically by the number of handles (genus) and the number
of boundaries.  Nonorientable surfaces are characterized additionally by the
number of crosscaps.  A crosscap is constructed by removing a circle from
the surface and identifying opposite points on the boundary.

The familiar M\"{o}bius strip can be constructed as a rectangle with one pair
of opposite edges identified in a nonorientable fashion.  This is similar to
the annulus, where the identification is orientable. We can make the single
boundary manifest by splitting the M\"{o}bius strip down the middle and
putting the two halves of the boundary next to each other (Fig.~1). 
We see that the M\"{o}bius strip has one boundary and one crosscap.  All
surfaces can be rearranged in this way so as to be shown to be constructed out
of boundaries, crosscaps and handles.  (For example, see~\cite{stillwell}.)

\Figure{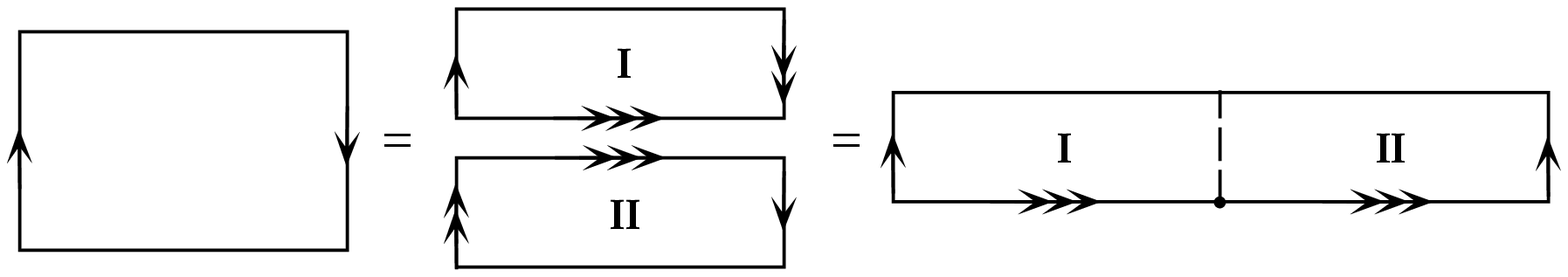 scaled 750}{Moving between realizations of the M\"{o}bius
strip.}

As the M\"{o}bius strip is to the annulus, the Klein bottle is to the torus;
the Klein bottle can be thought of as a cylinder with the opposite ends
identified nonorientably, or a rectangle with two opposite sides identified
orientably and the other pair, nonorientably.  Again, we can display the
surface in another form to make its crosscap structure manifest (Fig.~2); we
see that the Klein bottle is the surface with two crosscaps.

\Figure{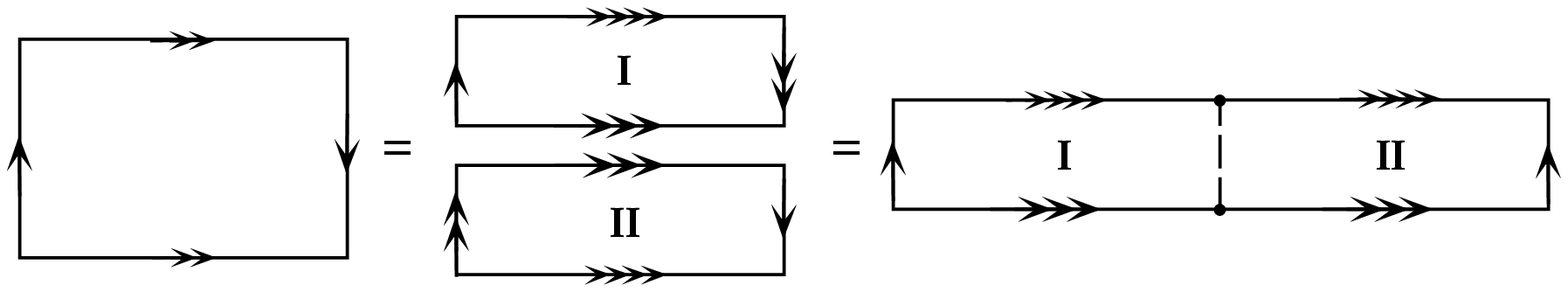 scaled 750}{Moving between realizations of the Klein bottle.}

While the torus has a complex conformal Killing vector --- corresponding to
the translation invariance of the torus inherited from the complex plane ---
the Klein bottle only has one real conformal Killing vector.  In the
two-crosscaps realization, this is easy to see; moving around the
circumference of the tube is a symmetry, but moving closer to or away from a
crosscap obviously is not.  In the cylinder realization, the symmetry is
translations along the tube, while moving around the circumference is not a
symmetry; this is because the nonorientable sewing induces two fixed points on
the ends of the cylinder.

The Klein bottle is also different from the torus in the structure of its
moduli space.  The torus has one complex modular parameter $\tau$, which
parameterizes the length of the cylinder and the angle through which it is
twisted before being sewn together.  The Klein bottle has just a real modulus,
corresponding to the length of the cylinder; a twist only returns another
equivalent Klein bottle.  Furthermore, the Klein bottle has no modular group.
The transformation $\tau \rightarrow - {1 \over \tau}$ produces
a distinct Klein bottle, because unlike the toral case where the two pairs
of sides are both identified in the same way, the two pairs of sides of the 
Klein bottle are identified in different ways.  The moduli space of the Klein
bottle is thus (0,$\infty$). 

For general surfaces, crosscaps, handles and boundaries all modify the 
Euler characteristic $\chi$: 

\begin{equation}
\chi = 2 - 2g - c - b - n -  {m \over 2}
\end{equation}

\noindent
where $g$ is the genus, $c$ the number of crosscaps, $b$ the number of
boundaries, $n$ the number of punctures on the surface and $m$ the number of
punctures on the boundary components.  This in turn dictates the number of
moduli of the surface

\begin{equation}
{\rm dim} \: {\cal M}^{g,c,n}_{b,m} = 6g - 6 + 2n + 3b + 3c + m.
\end{equation}

In all cases the number of crosscaps enters the formulae with half the
coefficient of the genus.  This is due to a theorem stating that in the
presence of a crosscap, a handle can always be decomposed into two more
crosscaps:~\cite{stillwell}

\begin{equation}
\rm handle + crosscap = 3 \ crosscaps.
\end{equation}

\noindent
Hence, when $c \neq 0$ the number of handles can be made zero and crosscaps, 
boundaries and punctures then specify the surface topologically.

In contrast to the case of oriented closed string theory, at each order
in perturbation theory several different moduli spaces contribute to a generic
amplitude in nonorientable open-closed string theory.  This occurs because
boundaries and crosscaps can also appear and contribute powers of the coupling,
resulting in different topologies at the same order of the expansion.  Looking
at string-scattering arguments, one can see that a crosscap contributes
$\kappa$, the gravitational coupling constant, to the order in perturbation
theory of a diagram.  One can then establish how many different moduli spaces
enter the computation of a given amplitude at each order.   

Since the contribution to the coupling from external states is the same for a
given amplitude at all orders, we can ignore this factor for the purposes of
seeing how many diagrams arise at each level.  Then we can concentrate on the
factor of the coupling $\kappa^l$ by which a given diagram differs from the
sphere with the same number of punctures.  A diagram with open-string punctures
has the additional requirement that there be at least one boundary.

Let us first look at oriented open-closed string theory.  Any diagram
contributing to the process at ${\cal O}(\kappa^l)$ must have a genus $g$ and
$b$ boundary components satisfying

\begin{equation}
l = 2g + b.
\end{equation}

\noindent
For external closed string states only, ${l+2 \over 2}$ moduli spaces
contribute for $l$ even and ${l+1 \over 2}$ for $l$ odd.  If there is at least 
one external open string, $l$ even has only $l \over 2$ distinct moduli spaces,
since we must have $b \neq 0$.  The result for $l$ odd is unchanged.

For the nonorientable sector, we can use the relation between crosscaps and
handles to characterize the surface solely in terms of crosscaps $c$,
boundaries $b$ and punctures.  Then diagrams of ${\cal O}(\kappa^l)$ have

\begin{equation}
l = c + b
\end{equation}

\noindent
and there are $l$ distinct moduli spaces at this order, where we have not
included $c=0$ as this falls in the orientable sector.  Hence there
are $l$ possibilities if we can have $b=0$ and $l-1$ if $b \neq 0$.  For the
nonorientable theory in general, both sectors contribute. 

\section{The Minimal Area Metric}

The length conditions of the minimal area problem for orientable closed and
open strings are that nontrivial closed curves must have at least length
$2 \pi$ and nontrivial open curves must have length at least
$\pi$~\cite{zwiebach}.  The first number is arbitrary and is chosen to
correspond to the circumference of the semiinfinite cylinders that represent
external closed strings.  The ratio of the values of the two conditions is a
variable parameter that defines a one-parameter family of string field
theories; the choice above corresponds to a theory where every open/closed
string diagram has as its double a closed string diagram~\cite{interpolate}.
For nonorientable surfaces, these length conditions are no longer adequate,
due to the appearance of a new class of closed curve, the ``M\"{o}bius''
curve.  We will examine how this curve arises, and we will be led naturally to
the new length condition that must be imposed.

Consider a closed curve $\gamma$ and consider a neighborhood of $\gamma$ along 
its length --- define two curves on either side of $\gamma$ that are locally
parallel to it by taking the set of all points a fixed distance along the
normal vectors to $\gamma$, and consider all the points between these two
flanking curves.  We can think of the curve as having been ``fattened'' into a
two-dimensional submanifold with boundary.  If this fattened curve is
topologically a M\"{o}bius strip, then $\gamma$ is a M\"{o}bius curve.  If, on
the other hand, the fattened curve is a simple annulus, the curve is the
ordinary orientable type we are used to, which we shall call an annular curve.

To see how the curves differ, let us examine the M\"{o}bius strip in
both of the realizations discussed previously.  First, consider the classic
realization of the M\"{o}bius strip as a strip that is twisted and attached to
itself.  A curve traveling along the strip parallel to the boundary will
generically loop around twice before returning to itself.  However, a curve
precisely at the center of the strip will come back to itself after only one 
loop (Fig.~3).  This curve, by definition, is a M\"{o}bius curve.  To see that
the other curve is annular, look at these same curves in the crosscap
realization.  Now the longer curve is seen to encircle the crosscap without
touching it --- it fattens to an annulus.  The M\"{o}bius curve is seen to
pass through the crosscap.  We see that the shortest M\"{o}bius curves are
just half the length of the shortest annular curves --- which interestingly
is the same length as the shortest nontrivial open curves.

\begin{figure} $$ \hSlide{-0.5in} \BoxedEPSF{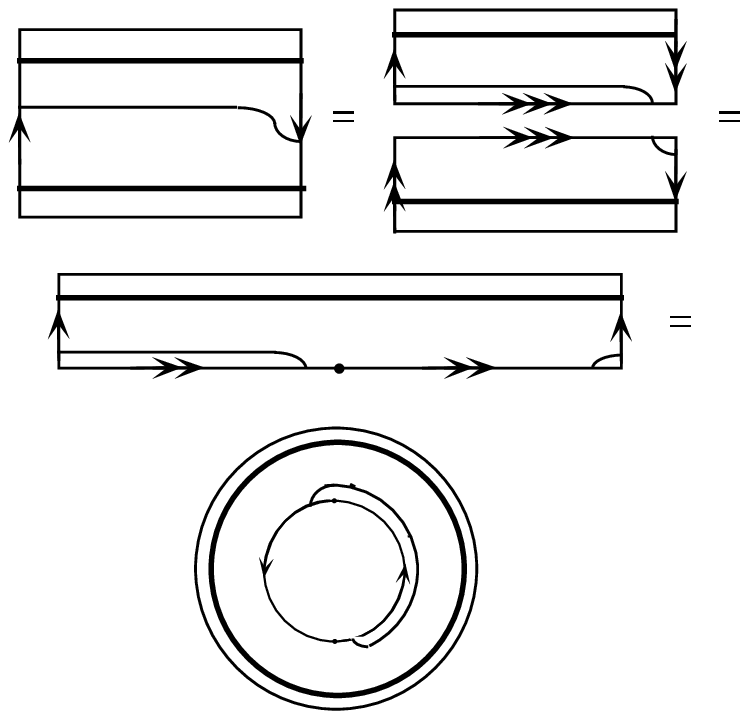}$$
\caption{\sf Two different classes of closed curve as seen on
the M\"{o}bius strip.  The heavy curve is a familiar annular curve, while the
lighter curve (deformed away from its minimal length for clarity) is the 
M\"{o}bius curve.}
\end{figure}

What would happen if we kept the same minimal area problem that was used for
orientable surfaces?  The closed-curve length conditions would apply to all 
nontrivial closed curves, M\"{o}bius and annular alike.  Consider
a cylinder ending on a crosscap.  In order to keep M\"{o}bius curves
going through the crosscap at least length $2 \pi$, the crosscap itself must
have circumference $4 \pi$.  However, the cylinder leading to the crosscap
need not have circumference any greater than $2 \pi$, and to minimize area it
will reduce to this size (Fig.~4).  Viewed in the M\"{o}bius strip
realization, the surface is more pathological.  There is no sensible way to
interpret a diagram such as this in terms of strings. 

\Figure{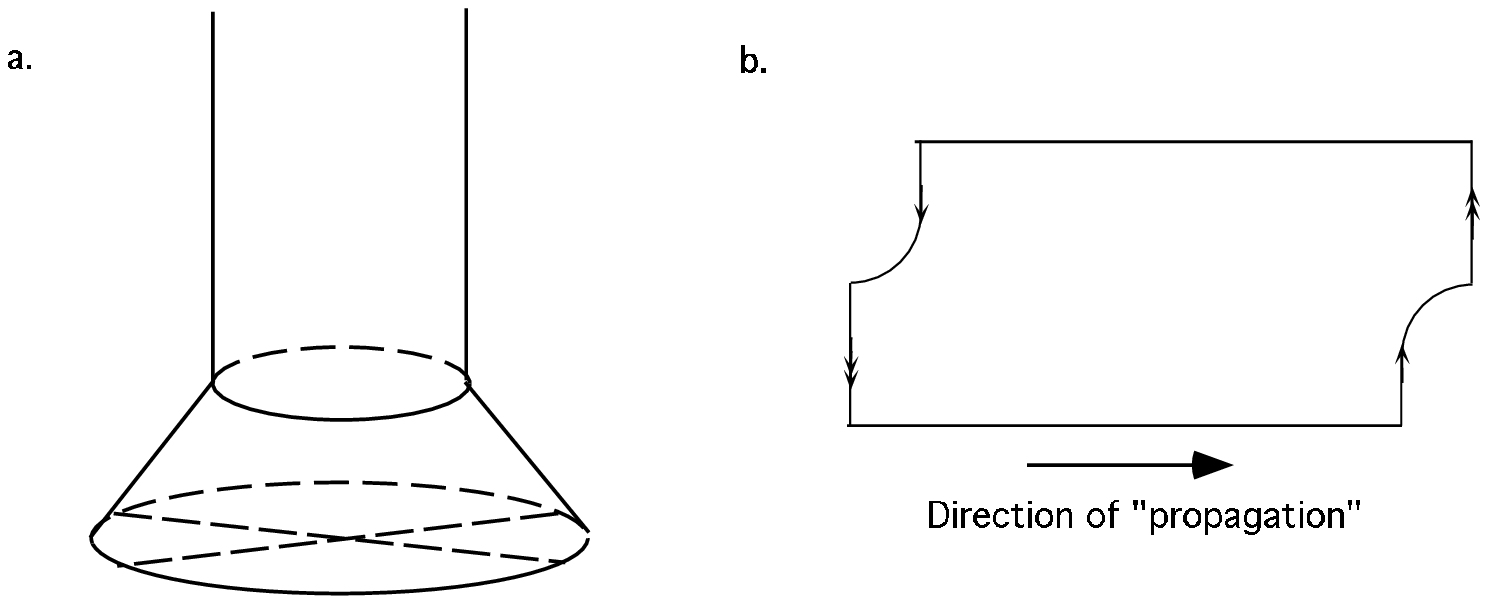 scaled 750}{A string ending in a crosscap when the naive
length conditions established for the orientable case are applied to all
curves on a nonorientable surface.  In a.~at left, the string must expand to
end in the crosscap; in b.~at right, in the alternate realization, there is no
interpretation in terms of propagating strings.}

Thus, the orientable length conditions are in need of modification for
the nonorientable case.  The natural thing to do is to require that nontrivial
annular closed curves have length $2 \pi$, while M\"{o}bius curves need only
have length $\pi$, and the same for the nontrivial open curves.  In fact, as 
we shall prove below, it is sufficient only to set conditions on the annular
closed and the open curves; the length conditions on the M\"{o}bius curves
follow.

To see this, consider a crosscap and a M\"{o}bius curve $\alpha_1$, wrapped
around half the crosscap, and thus as short as it can be.  Impose the
conditions  that nontrivial
annular curves have length of at least $2 \pi$, but impose no condition on the
M\"{o}bius curves.  Let $\alpha_1$ have a length less than $\pi$; we seek to
show that then there must exist a nontrivial annular curve with length less
than $2 \pi$, in contradiction with our assumptions.  There is another
M\"{o}bius curve, $\alpha_2$, which goes around the other half of the
crosscap.  It is clear that the smallest annular curve that can surround the
crosscap $\gamma$ has a length equal to the sum of the two M\"{o}bius curves.
If the metric on the $\alpha_2$ side of the crosscap is the same as
the metric at $\alpha_1$ --- in other words, if the metric is continuous
across the crosscap --- the two have the same length and their sum, and thus
the annular curve's length, is less than $2 \pi$, our desired contradiction.
Imagine that the is not continuous across the crosscap, and
thus that $\alpha_1$ has different (say, shorter) length than $\alpha_2$.  The
metrics on either side of the crosscap in general vary with position, but since
$l_{\alpha_1} < l_{\alpha_2}$, there must be at least one point on the 2 side 
with metric greater than at least one point on the 1 side.  We can lower the
metric on the 2 side at this point while raising it on the 1 side in such a
way as to preserve $l_{\gamma} = l_{\alpha_1} + l_{\alpha_2}$; the
perturbation is simply

\begin{eqnarray}
\rho_1 \rightarrow \rho_1 + \delrho \\
\rho_2 \rightarrow \rho_2 - \delrho 
\end{eqnarray}

\noindent
where $ ds^2 = \rho^2 \, dz d\bar{z}$,
$\delta \rho > 0$ and the metric deformations take place in squares of
infinitesimal side length $\epsilon$ centered on the points in question.
The total area is quadratic in the metric, however, and changes as

\begin{equation}
{\cal A} \rightarrow {\cal A}' = {\cal A} + 2 \, \delrho \, \epsilon^2
(\rho_1 - \rho_2) + {\cal O}(\delrho^2) < {\cal A}.
\end{equation}

We have found a metric perturbation that preserves length conditions, but
lowers the total area.  Thus the metric configuration with a discontinuity
across a crosscap is not of minimal area.

We see placing conditions on the annular
curves is sufficient to induce the appropriate conditions on the M\"{o}bius
curves as well.  This is satisfying, since we need only place conditions on
curves that correspond to propagators and external string states.

We are now prepared to state the minimal area problem for nonorientable
open/closed strings:

\goodbreak
\vskip0.25in
\noindent
\bf Minimal Area Problem for Nonorientable Open-Closed String Theory: \rm  
Given a genus $g$ surface $R$ with $b$ boundaries, $c$ crosscaps, $m$
punctures on the boundaries and $n$ punctures in the interior ($g,b,c,n,m \geq
0$) the string diagram is defined by the metric of minimal (reduced) area under
the condition that the length of any nontrivial open curve in $R$ with 
endpoints at the boundaries be greater than or equal to $\pi$ and the length
of any nontrivial {\it annular} closed curve be greater than or equal to $2
\pi$.

\section{Two explicit examples}

Diagrams are constructed from vertices and propagators, sewing together
surfaces with fewer moduli to span part of moduli space, and covering the
rest by adding the necessary diagrams to that moduli space's vertex.
To illuminate the methods, we will construct the diagrams for two different
cases.  First we will examine the Klein bottle, which is interesting as a 
nonorientable version of the torus.  Then, for a more complicated example,
we will look at the Klein bottle with boundary, a surface with two crosscaps
and one boundary.

\vskip.1in
\noindent
{\bf The Klein Bottle}
\vskip.05in

As we have discussed, the Klein bottle is a zero-genus surface with two
crosscaps and no boundaries.  We begin by constructing all possible Klein
bottles from more elementary diagrams and seeing how much of moduli space we
span this way.  We can self-sew arbitrarily long closed-string propagators
nonorientably, and thus obtain Klein bottles at various points in moduli space
all the way to infinity (Fig.~5a).  However, we cannot make the propagator too
short, for if its length is less than $\pi$, the M\"{o}bius closed curve
around its middle will violate the length conditions.  It becomes necessary for
us to go to another ``channel'' to span the part of moduli space with small
values of the parameter.  

The other way to make a surface with two crosscaps is to sew together two
closed string self-interaction vertices, with an arbitrarily long closed
string propagator in the middle (Fig.~5b).  Viewing this two-crosscaps
realization in the more familiar self-sewn cylinder realization, we see that
very long two-crosscap diagrams correspond to cylinders of length $\pi$ 
(satisfying length conditions) but with arbitrarily large circumferences.
Since it is the ratio of length to circumference that matters, these
correspond to arbitrarily small values of the modular parameter.

Thus we have filled in both ends of moduli space.  Additionally, we see
that the two channels come together smoothly in the middle, at the surface
which is a cylinder of length $\pi$ and circumference $2 \pi$, saturating
the length conditions (Fig.~5c).  We have
now spanned moduli space for the Klein bottle successfully.

\Figure{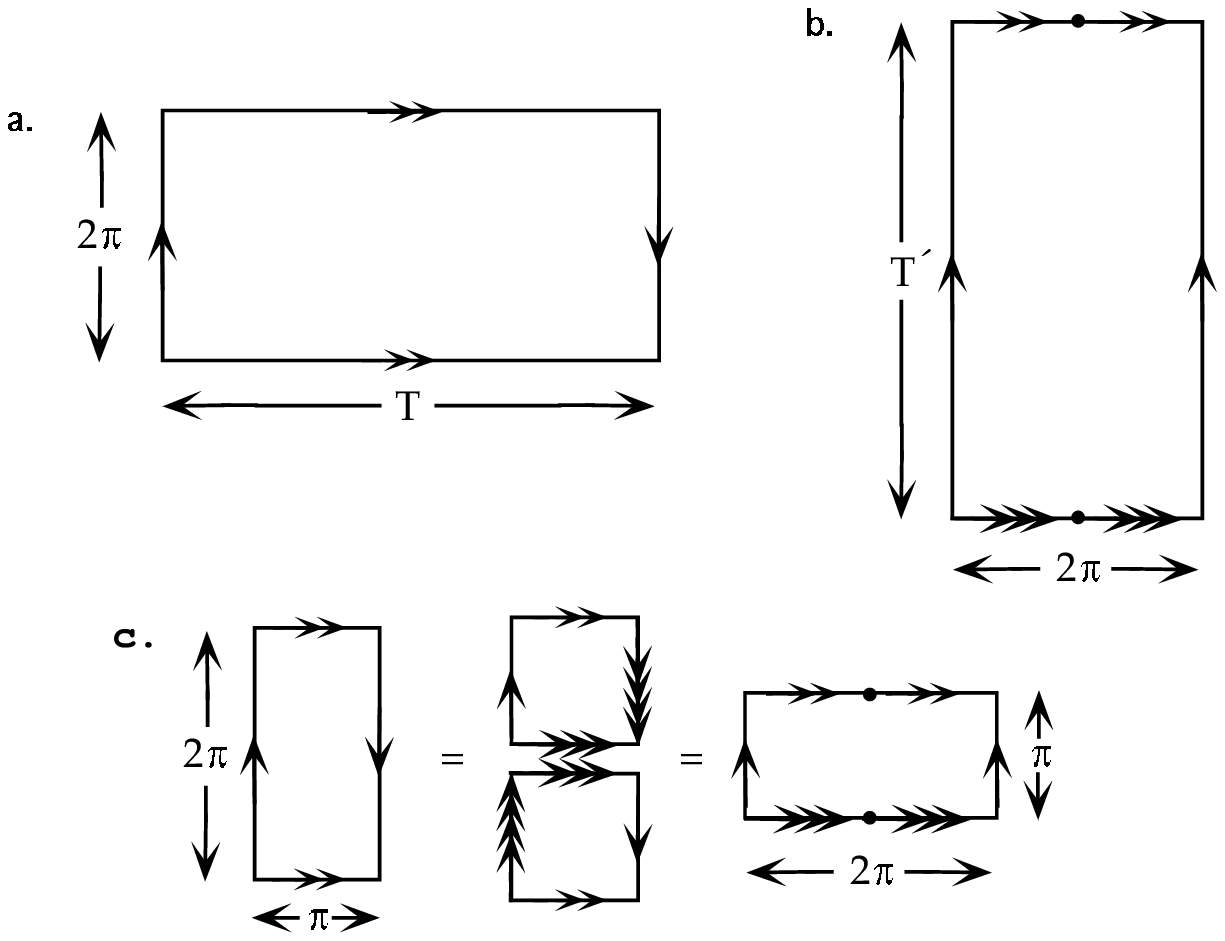 scaled 750}{Spanning the moduli space of the Klein bottle.  Large 
values of the modular parameter are obtained by self-sewing an arbitrarily
long propagator nonorientably, as in a.  Small values are achieved in the
other realization of the Klein bottle, by connecting two crosscaps with a
propagator of arbitrary length, as in b.  The two channels come together
smoothly when the surfaces saturate their length conditions, at $T = T' = \pi,$
as shown in c.}

\vskip.25in
\noindent
{\bf Klein bottle with boundary}
\vskip.1in

This is a considerably more complicated surface, and we will not describe the
process for spanning moduli space in quite the same detail.  There are three
real moduli.  These can be thought of as

\vskip.1in

\begin{enumerate}
\item the length of the Klein bottle

\item the size of the boundary, and

\item the location of the boundary.
\end{enumerate}

\vskip.05in

\noindent
The location of the boundary is one real parameter because the Klein bottle
has one real conformal Killing vector, rendering the location of the boundary
in one of the two dimensions of the surface irrelevant.  The location in
the other direction is relevant; in fact, movement along this direction is
equivalent to a twist of the bottle.

\Figure{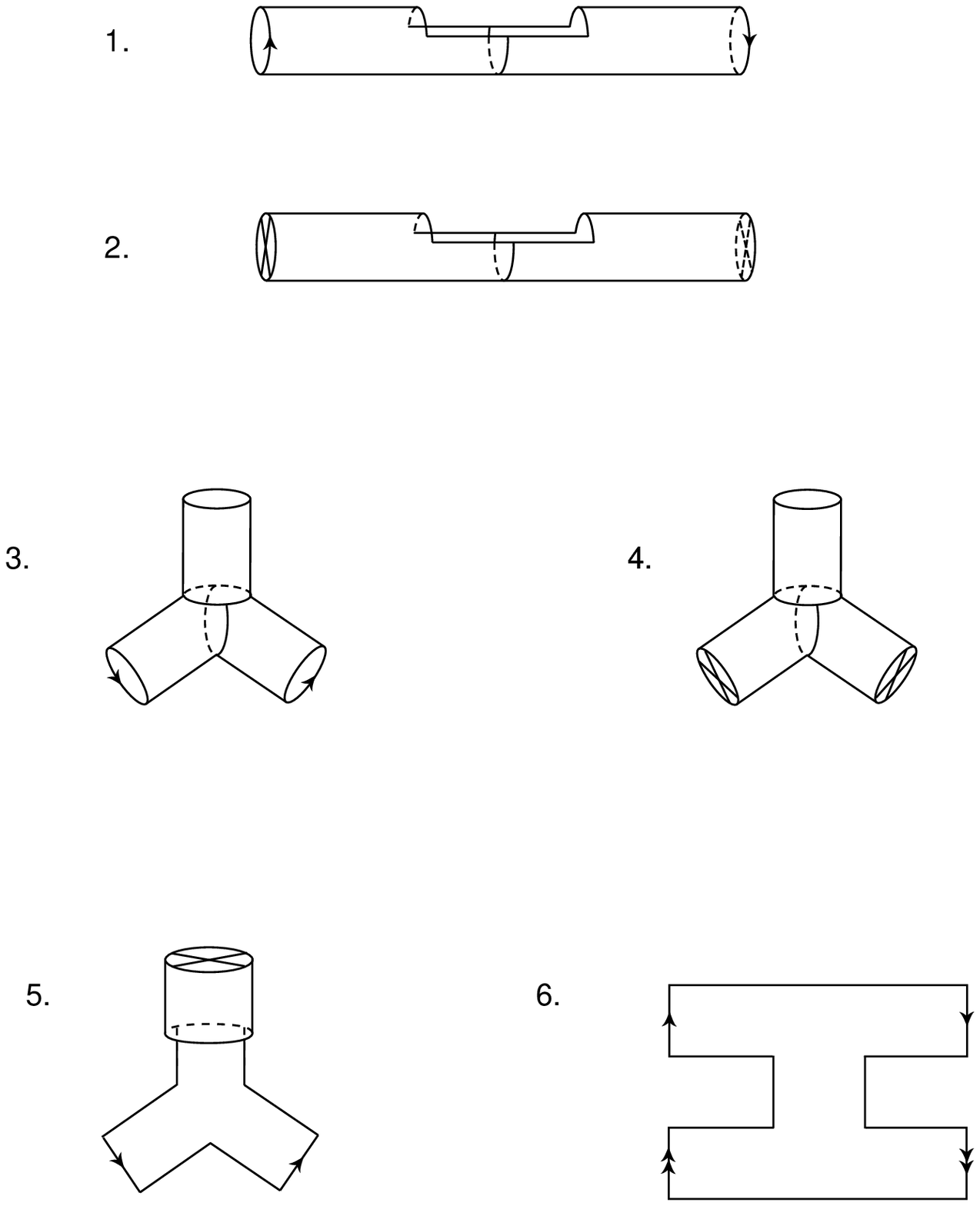 scaled 750}{The six ways of constructing the Klein bottle with
boundary .}

There are six ways of constructing the surface with two crosscaps and a
boundary; they are displayed in Figure 6.  It is not hard to show that
together they fill in the ``corners'' of moduli space, that is, that the
asymptotic regions of moduli space can be covered by allowing the various
propagators to get very large.  Once again what is large in one channel
corresponds to a small value of moduli in another.

It is a little bit of work to show that the six all come together smoothly in
the middle.  The large-boundary and small-boundary diagrams (1 and 3, and 2 and
4, respectively) interpolate into one another smoothly as shown in Figure 7. 
To connect the first and second diagrams, for example, we look at them both in
the same realization and see what happens to the boundary when the Klein bottle
saturates its length conditions (Fig.~8).  It is most naturally reduced to a
slit in each case, but the slits are at right angles to each other and in
different locations.  The location of one slit can be shifted to the other by
twisting the Klein bottle.  In order for the boundaries to interpolate into
each other, we must have instead of the naive slit a progression through a
square as shown in Figure 9.  This connects the first four diagrams; the last
two can be done similarly.

\Figure{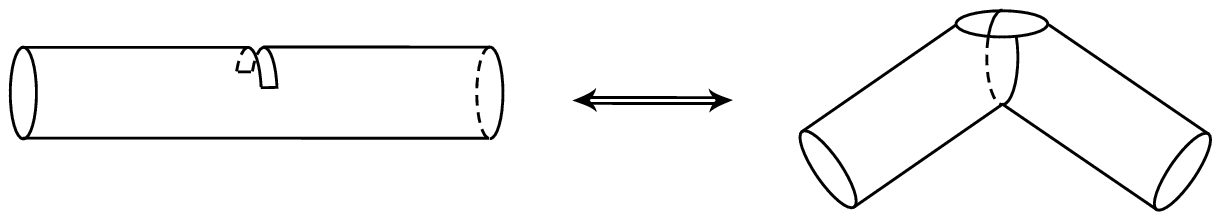 scaled 750}{The interpolation between large- and small-boundary diagrams.}

\Figure{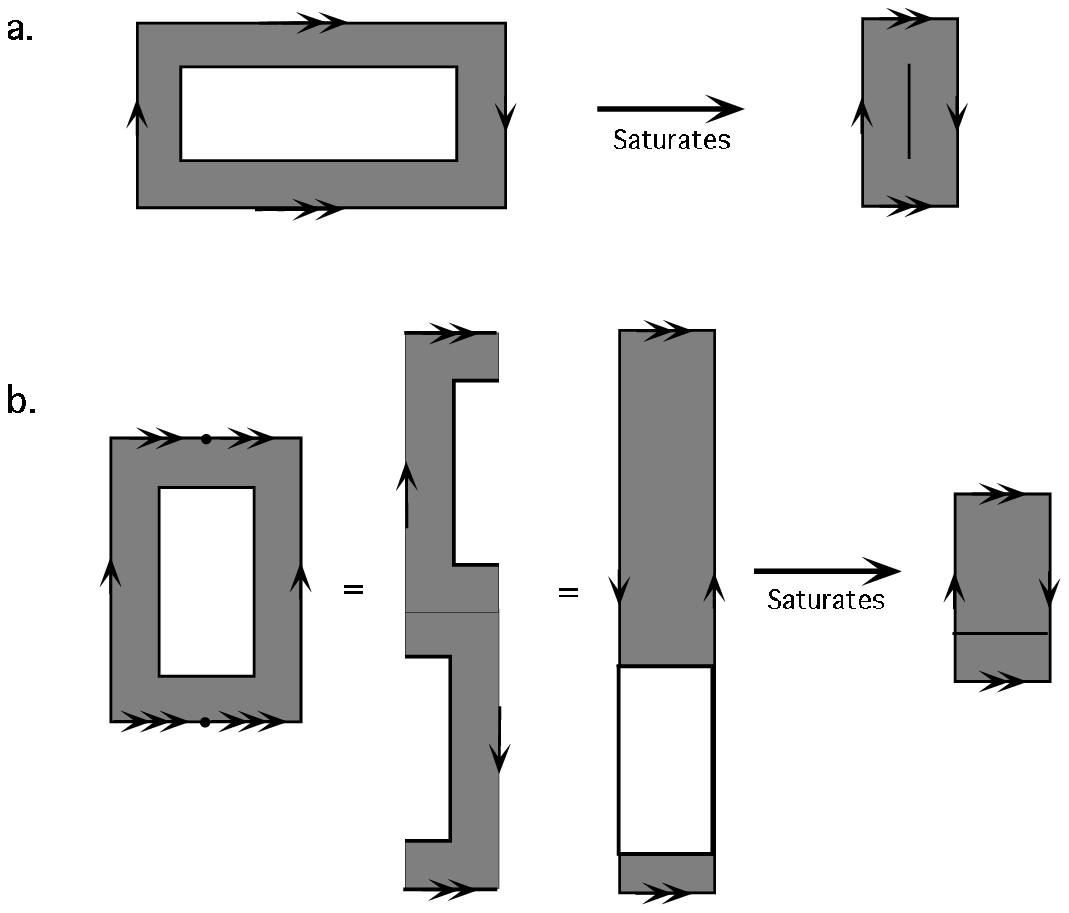}{The surfaces identified as 1 and 2 on Figure 6 saturating
the length conditions on the Klein bottle.  The surfaces are shaded to keep
the location of the boundary clear.}
\Figure{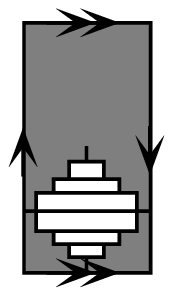}{The actual evolution of the boundary as surfaces
1 and 2 come together.}

\newpage

\section{Sewing Theorem and Sewing Recursion Relations}

\noindent
{\bf The Sewing Theorem}
\vskip.1in

Here we generalize the sewing theorem of Wolf and Zwiebach~\cite{wolf}.  One
wants to assure that when two surfaces which have minimal area metrics
satisfying the length conditions (``admissible'') are sewn together, or when a 
single surface is sewn to itself, the resulting surface also has an admissible 
minimal area metric.  Thus we must ascertain that after the process of sewing
there are no nontrivial curves which violate the length conditions.  The result
turns out to be a straightforward extension of the sewing theorem for
orientable surfaces.  There are three cases to consider.

First consider the curves which do not pass through the sewing line (the line
$|z|=|w|=1$ when sewing with $zw=1$), and which were nontrivial prior to
sewing.  The process of sewing only 
involves two steps: amputating the semiinfinite cylinders/strips that
represent legs which are to be sewn, and then identifying the stubs.  But the
metric used after amputation $\rho_0$ is just the restriction of the original
metric $\rho$ to the amputated surface; hence all the curves on the surface
retain the same length.  Since sewing is then simply a process of
identification, which does not affect the lengths of curves, we see that all
curves that satisfy the length conditions before sewing, continue to do so
after sewing, regardless of type.

Now we must consider the curves which were trivial before sewing.  These also
retain the same length, but if a short one were to become nontrivial in the
sewing process, the length conditions would be violated.  A trivial closed
curve $\gamma$ must bound a disc $D$, regardless of whether the surface is
orientable or nonorientable.  (The proof uses the universal covering space,
and is not modified much for the nonorientable case, since every nonorientable
two-dimensional surface has an orientable surface as its
cover~\cite{stillwell}.  For the proof on Riemann surfaces,
see~\cite{strebel}.)  However
the process of sewing does not affect this disc; sewing only connects two
punctures, and since $D$ cannot contain a puncture and still be a disc, it
remains a disc after the sewing.  Thus $\gamma$ remains trivial.  A trivial
open curve also can be seen to bound half a disc.  Any trivial open 
curve must begin and end on the same boundary component.  If the points on the
boundary where the curve $\beta$ begins and ends are $A$ and $B$, then $\beta$ 
and line segment $\overline{AB}$ must bound a disc.  The same argument as above
then shows that the curve must remain trivial.
There are no trivial M\"{o}bius curves;  a trivial curve must bound a disc,
but if we ``fatten'' up a curve bounding a disc, we obtain an annulus, not a
M\"{o}bius strip.  

Now we examine the case of curves that pass through the seam.  A
curve that is homotopic to the seam must at least be the length of the seam
itself, and the seam satisfies the length conditions, so these curves are safe.
For curves not homotopic to the seam, we have
used the choice of where to amputate the semiinfinite cylinders and
strips to our advantage.  We deliberately keep stubs of length $\pi$ to 
ensure that the length conditions are satisfied in the new surface.  The
curves must extend
out of the cylinder/strip and into the rest of the surface, where there are
other punctures, boundary components, cycles etc. which are necessary to make
them nontrivial.  Thus the curve must traverse either one of the stubs
twice or each stub once, acquiring thereby a length of no less than $2 \pi$.
Thus it satisfies the length conditions regardless of the type of curve. 

Hence the sewing of two surfaces with admissible metrics always produces
another surface with an admissible metric, and the sewing theorem is complete.

\vskip.15in

\noindent
{\bf Sewing Recursion Relations}
\vskip.1in

Different classes of diagrams cover different regions of moduli space, and
for the complete covering to be sensible, recursion relations must hold
at the boundaries of the different regions.  Specifically, the vertex covers 
a certain region of moduli space; the string diagrams on the boundary of
this region must match with diagrams created from more elementary vertices
with one collapsed propagator.  This is expressed by the master equation

\begin{equation}
\del {\V} + \frac{1}{2}\{ {\V},{\V} \} + \hbar \Delta {\V} = 0
\end{equation}

\noindent
where the antibracket operation $ \{\ ,\ \}$ sews together two different
surfaces and $\Delta$ sews a single surface to itself, and the full vertex
${\V}$ will be defined below.  For details, see~\cite{revisited}.  Each 
of these operations contains all possible relevant types of sewing, closed
and open.  In considering the nonorientable case,
one has an additional new way to sew, nonorientable sewing.

When sewing two generic surfaces with coordinates $z$ and $w$ in the
neighborhoods of the punctures to be connected, one sews by identifying the
coordinates:  $zw = {\it const}$.  However nonorientability permits a new kind
of identification: $z\overline{w} = {\it const}$.  This ``antiholomorphic''
identification
produces a nonorientable surface when an orientable surface is self-sewn.
Holomorphic and antiholomorphic coordinates on the same surface are linked;
orientability is thus destroyed, as a small oriented circle (indicatrix) can
pass through the seam along a closed curve and return with its orientation
reversed.

Consider sewing two disjoint surfaces in an antiholomorphic fashion.  Sewing
surface $A$
to surface $B$ in this way is equivalent to holomorphically sewing surface $A$
to the mirror of surface $B$, that is, the surface resulting from exchanging
the holomorphic and antiholomorphic coordinates on $B$.  But the mirror of $B$
is just another surface, with the same values of $g$, $b$, $c$, etc.~as $B$
itself, and already present in the collection of surfaces under consideration.
Thus to sew two disconnected surfaces antiholomorphically produces the same
resulting surface as a different holomorphic sewing.  Hence if all surfaces
are sewn both holomorphically and antiholomorphically, the antibracket
operation will produce each surface exactly twice.

The propagator is fixed, regardless of which surfaces are being sewn.  Hence we
require a single prescription incorporating both kinds of sewing:

\begin{equation}
{\rm Propagator} \equiv \frac{1}{2} ({\it sew} + {\it \overline{sew}})
\end{equation}

\noindent
where {\it sew} and ${\it \overline{sew}}$ denote holomorphic and antiholomorphic sewing,
respectively.  The factor of $\frac{1}{2}$ ensures that this prescription
will produce each surface once when different surfaces are sewn via the
$ \{\ ,\ \}$ operation.  The $\Delta$ operation will then produce two different
resulting surfaces for each surface it is applied to, with the factor of
$\frac{1}{2}$.

Sewing operations change the topology of the surface(s) on which they are 
performed.  The orientable sewing operations have been enumerated in~\cite{revisited}.  Antiholomorphic sewing of different surfaces has precisely the 
same topological consequence as its holomorphic counterpart.  Antiholomorphic
sewing of the same surface produces crosscaps, unique to the nonorientable
case.  These operations are summarized in Table 1, for a surface with genus
$g$, $b$ boundaries, $c$ crosscaps, and $n$ closed and $m$ open punctures.

\begin{table}
\begin{tabular}{||c|c|c|c|c||} \hline \hline
String type sewn & Surfaces Sewn & Boundaries Sewn & Orientation & Change in Topology \\ \hline \hline
  Closed  & Different & --- & {\it sew}, ${\it \overline{sew}}$ & --- \\ \hline
  Closed  & Same      & --- & {\it sew}  & $g \ra g+1$ \\ \hline 
 Closed &  Same   & --- & ${\it \overline{sew}}$ & $c \ra c+2$ \\ \hline
 Open &  Different & Different & {\it sew}, ${\it \overline{sew}}$ & $b \ra b-1$ \\ \hline
 Open &  Same  &   Same  &  {\it sew} & $b \ra b +1$ \\ \hline
 Open & Same &  Same & ${\it \overline{sew}}$ &  $c \ra c+1$ \\ \hline
  Open  & Same & Different & {\it sew} & \( \begin{array}{c}
	                                                   g \ra g+1 \\
                                                           b \ra b-1 \\
                                                     \end{array} \) \\ \hline
 Open & Same & Different & ${\it \overline{sew}}$ & \( \begin{array}{c}
                                                                   c \ra c+2 \\
                                                                   b \ra b-1 \\
                                                              \end{array} \) \\
\hline \hline
\end{tabular}
\caption{\sf How each distinct sewing operation affects the topology of
the resulting surface.  For sewing of different surfaces, $g$ denotes $g_1 +
g_2$, etc.}
\end{table}

The sewing recursion relations on a vertex from a single moduli space do
not involve the full form of the propagator, which
includes both kinds of sewing on an equal footing.  For the $\Delta$ operation
the two types of sewing change the topology
of the surface in different ways, and hence on the boundary of a given vertex
it is a different surface which appears with a holomorphically sewn collapsed
propagator than the one that appears with an antiholomorphically collapsed
propagator.

One can remedy this situation by collecting the various vertices 
corresponding to surfaces with fixed values of $n$ and $m$ and a given Euler
characteristic (or equivalently, the various vertices of a given amplitude
at a certain order of the coupling constant) into a single vertex
${\V}^{\overline{\chi}}_{n,m}$:

\begin{equation}
{\V}^{\overline{\chi}}_{n,m} = \sum_{g,c,b}^{'} {\V}^{g,c,b}_{n,m}
\end{equation}

\noindent 
where the summation is constrained to sum over all possible values of $g,c,b$
that produce the appropriate value of\ $\overline{\chi}$, the Euler
characteristic of the surface composing the vertex.  The recursion relations
applied to the objects ${\V}^{\overline{\chi}}_{n,m}$ involve the full
propagator: although the two different kinds of sewing change the topology in
different ways, they modify the Euler characteristic of the surface by the same
amount, as one can verify with the help of Table 1. 

The total vertex with appropriate values of the coupling $\kappa$ and $\hbar$
is then:

\begin{equation}
{\V} \equiv \sum_{n,m,\overline{\chi}} \hbar^{-\overline{\chi} + \overline{p}} \kappa^{-2 \overline{\chi}}\ {\V}^{\overline{\chi}}_{n,m}
\end{equation}

\noindent
where $\overline{p} = 1 - \frac{1}{2}(n+m)$~\cite{revisited}.  Implicit in the
sum over open punctures $m$ is a sum over all possible 
distinct distributions of the punctures over the various boundary components.
This vertex then satisfies the master equation, Eq.\ 9.

\vskip.3in

\subsection*{Acknowledgments}

I would like to express my gratitude to Barton Zwiebach, who
proposed this problem and provided guidance.  Also, I would like to thank
Rachel Cohen and Marty Stock for help with the figures, and Benjamin Scarlet
for useful discussions and \LaTeX\ troubleshooting.

\end{document}